\documentclass[a4paper,fleqn,usenatbib]{mnras}

\usepackage[T1]{fontenc}
\usepackage{ae,aecompl}

\usepackage{graphicx}	
\usepackage{amsmath}	
\usepackage{amssymb}	
\usepackage{epstopdf}
\usepackage{array}
\usepackage{multirow}
 
\newcommand\MyBox[2]{
  \fbox{\lower0.75cm
    \vbox to 1.7cm{\vfil
      \hbox to 1.7cm{\hfil\parbox{1.4cm}{#1\\#2}\hfil}
      \vfil}%
  }%
} 
 
\title[AstroML]{Classifying Complex Faraday Spectra with Convolutional Neural Networks}
\author[Brown et al. 2017]{
Shea Brown$^{1}$\thanks{E-mail:shea-brown@uiowa.edu},  
Brandon Bergerud$^{1}$, 
Allison Costa$^{1}$,
B. M. Gaensler$^{2}$,\newauthor
Jacob Isbell$^{1}$, 
Daniel LaRocca$^{1}$,
Ray Norris$^{3}$, 
Cormac Purcell$^{4}$, \newauthor
Lawrence Rudnick$^{5}$,
Xiaohui Sun$^{6}$
\\
$^{1}$Department of Physics \& Astronomy, The University of Iowa, Iowa City, IA, 52245\\
$^{2}$Dunlap Institute for Astronomy and Astrophysics, The University of Toronto, Toronto, ON M5S 3H4, Canada \\
$^{3}$Western Sydney University, Locked Bag 1797, 1797, Penrith South, NSW, Australia \\
$^{4}$Research Centre for Astronomy, Astrophysics, and Astrophotonics, Macquarie University, NSW 2109, Australia \\
$^{5}$Minnesota Institute for Astrophysics, University of Minnesota, 116 Church Street SE, Minneapolis, MN 55455, USA\\
$^{6}$Department of Astronomy, Yunnan University, and Key Laboratory of Astroparticle Physics of Yunnan Province, Kunming, 650091, China
}

\begin{document}

\date{}

\maketitle

\label{firstpage}

\begin{abstract}  Advances in radio spectro-polarimetry offer the possibility to disentangle complex regions where relativistic and thermal plasmas mix in the interstellar and intergalactic media. Recent work has shown that apparently simple Faraday Rotation Measure (RM) spectra can be generated by complex sources. This is true even when the distribution of RMs in the complex source greatly exceeds the errors associated with a single component fit to the peak of the Faraday spectrum. We present a convolutional neural network (CNN) that can differentiate between simple Faraday thin spectra and those that contain multiple or Faraday thick sources. We demonstrate that this CNN, trained for the upcoming Polarisation Sky Survey of the Universe's Magnetism (POSSUM) early science observations, can identify two component sources 99\% of the time, provided that the sources are separated in Faraday depth by $>$10\% of the FWHM of the Faraday Point Spread Function, the polarized flux ratio of the sources is  $>$0.1, and that the Signal-to-Noise radio (S/N) of the primary component is $>$5. With this S/N cut-off, the false positive rate (simple sources mis-classified as complex) is  $<$0.3\%. Work is ongoing to include Faraday thick sources in the training and testing of the CNN. 
\end{abstract}


\section{Introduction} 

\subsection{Background}
Faraday rotation of linearly polarized radio emission gives unique insight into the properties of the intervening magneto-ionic medium. Measurements of rotation-measures of background polarized radio sources probes  astrophysical magnetic fields in a variety of environments like the Solar corona \citep{kooi17}, H~{\scriptsize II} regions \citep{harv11}, the interstellar medium of the Milky Way \citep{han97, sun10, woll10, pshi11, vane11, jans12, akah13}, external galaxies \citep{han98, gaen05, mao12, bern13}, and the intracluster \citep{bona10,bona13} and intergalactic \citep{akah11, akah14} medium. Traditionally, Faraday rotation has been measured by fitting the change in polarization angle $\chi$ as a function of wavelength squared ($\lambda^2$), parameterized by the rotation measure (RM) defined by 

\begin{equation}\label{eq:chi} 
\chi (\lambda^2 ) = \chi_0 + RM\lambda^2 , 
\end{equation}

\noindent where $\chi_0$ is the intrinsic polarisation angle of the radio emission. This linearity with $\lambda^2$ is only valid for the case of a single synchrotron emitting source with an intervening cloud of magnetised thermal plasma.   The wide-band capability of modern radio telescopes has allowed the use of RM synthesis \citep{bren05, sun15}, which can address problems of bandwidth depolarization and multiple emitting/rotating regions along the line of sight (or within the same beam).  RM Synthesis inverts the complex polarisation spectrum $P(\lambda^2) = Q(\lambda^2) + i U(\lambda^2)$ into a Faraday spectrum 

\begin{equation}
F(\phi)  =  K \int_{-\infty}^{+\infty}P
(\lambda^2)\mathrm{e}^{-2\mathrm{i}\phi (\lambda^2-\lambda_0^2)}\ 
\mathrm{d}\lambda^2 \label{inversion}
\end{equation}

\noindent where $K$ is a constant, and $\phi$ is the ``Faraday depth" of the emission, given by
\begin{equation}\label{eq:defF}
\phi(r) = K'\int_{\vec{r}}^{0} n_e\vec{B}\cdot\vec{{\rm d}l}, 
\end{equation}
\noindent where $K'$ is a constant, $n_e$ is the electron density, $\vec{B}$ is the magnetic field vector, and $\vec{{\rm d}l}$ is a infinitesimal distance along the line of sight from the synchrotron source located at $\vec{r}$. In the simple case described in Equation \ref{eq:chi}, $\phi$=RM. The development of deconvolution algorithms for the Faraday spectra \citep{heal09} has further improved the ability of RM Synthesis to reveal multiple sources along the line of sight.

\cite{farn11} described an ambiguity in RM derived from $\chi(\lambda^{2})$ fitting and RM Synthesis where two Faraday component model can produce a consistent single component solution that is neither of the input components nor their mean. This ambiguity can lead to an error in $\phi$ derived from these methods that is greater than what one would naively calculate from the uncertainty in fitting the peak of $F(\phi)$. In the era of large radio surveys meant to produce grids of background RMs for the archival science, there is  a need to distinguish between ``simple" foreground screens and more complex sources.  


One such survey is the Polarisation Sky Survey of the Universe's Magnetism \citep[POSSUM,][]{gaen10}, which will be conducted with the Australian Square Kilometre Array Pathfinder \citep[ASKAP,][]{john08} and will measure more than 1 million polarized sources in the frequency range of 1130-1430~MHz over 75\% of the sky. An Early Science survey, which is being conducted as part of ASKAP's science commissioning observations, will make use of only 12 antennae of ASKAP, but the frequency coverage will be extended to 700-1800~MHz. The extended frequency coverage of the Early Science survey is ideal for identifying and investigating complex Faraday spectra, provided that these spectra can be identified in an automated way. There are a variety of ways in which a Faraday spectrum can deviate from a single source with a foreground Faraday screen \citep[called Faraday thin,][]{bren05}. It can be Faraday thick (caused by significant mixing of Faraday rotating {\it and} emitting plasma), have multiple Faraday thin components, or there can be external/internal Faraday dispersion (modulation due to rapidly changing Faraday rotating cells along the line of sight, or within a single beam).

The initial data release of the POSSUM survey will be a catalogue of sources with {\it simple} Faraday spectra and their associated properties. Simple spectra come from sources with a single Faraday rotating screen in front of them, i.e., their polarisation angle would obey Equation 1 for all $\lambda^2$ values. This POSSUM Polarization Catalogue (PPC) will not include complex sources, which are any sources that are not simple as defined above, so the pipeline producing the catalogue must have an efficient and effective way of determining/testing the complexity of a source. This test needs to be able to: 1) determine whether a Faraday spectrum is complex in general, including cases where there are more than one well separated Faraday sources (peaks) in the spectrum, 2) determine whether a given peak is Faraday thin or not, and 3) provide some way of assessing how sure we are of the resulting classification. Initial work has shown that the second moment of the clean components resulting from a Faraday cleaning procedure can provide some discriminating power \citep[e.g.,][]{brow11,ande15}, while more recent work has focused on the model fitting the polarization spectrum and examining the statistically more likely model in a Bayesian sense \citep[e.g.,][]{osul12,sun15,osul17,purc17}.

To this end, we present the construction of a convolutional neural network (CNN) that can classify a Faraday spectrum as either simple or complex. In $\S$2 we describe the construction and training of the CNN, in $\S$3 we outline the testing of the network on simulated data, and in $\S$4 we summarize the limitations of the metric and discuss future improvements. 

\section{Convolutional Neural Network}
A few of the major difficulties for developing a test for complexity are A) the ad hoc nature of the initial choice of a metric, B) the significant work required to find a metric threshold appropriate for the science goals, and C) estimating the uncertainty in the accuracy of the metric/threshold combination. For these reasons, we have explored convolutional neural networks \citep[CNNs,][]{lecu95,kriz12}, in particular the ``inception" networks developed by the GoogLeNet team \citep{szeg14} as potential classifiers. CNNs are ideal for this problem, as they apply a series of convolutions to the data, along with non-linear intermediate functions \citep[e.g., rectified linear units, or ReLUs,][]{nair10}, and the training process will find the series of convolutional kernels that are optimal for classifying the spectra. While model fitting requires large searches over parameter space each time a spectrum is analyzed, a CNN samples the space only during training, and requires only a straightforward and efficient feed-forward network of matrix operations to classify the sources. In the inception model a series of convolutions is done in parallel, each with a different kernel size, further allowing the network to search for features in the data that distinguish complex sources.  In essence, the network will find the best metric to use during the training process, eliminating the need to choose one at the outset. 

\begin{figure}
\begin{center}
\includegraphics[width=8cm]{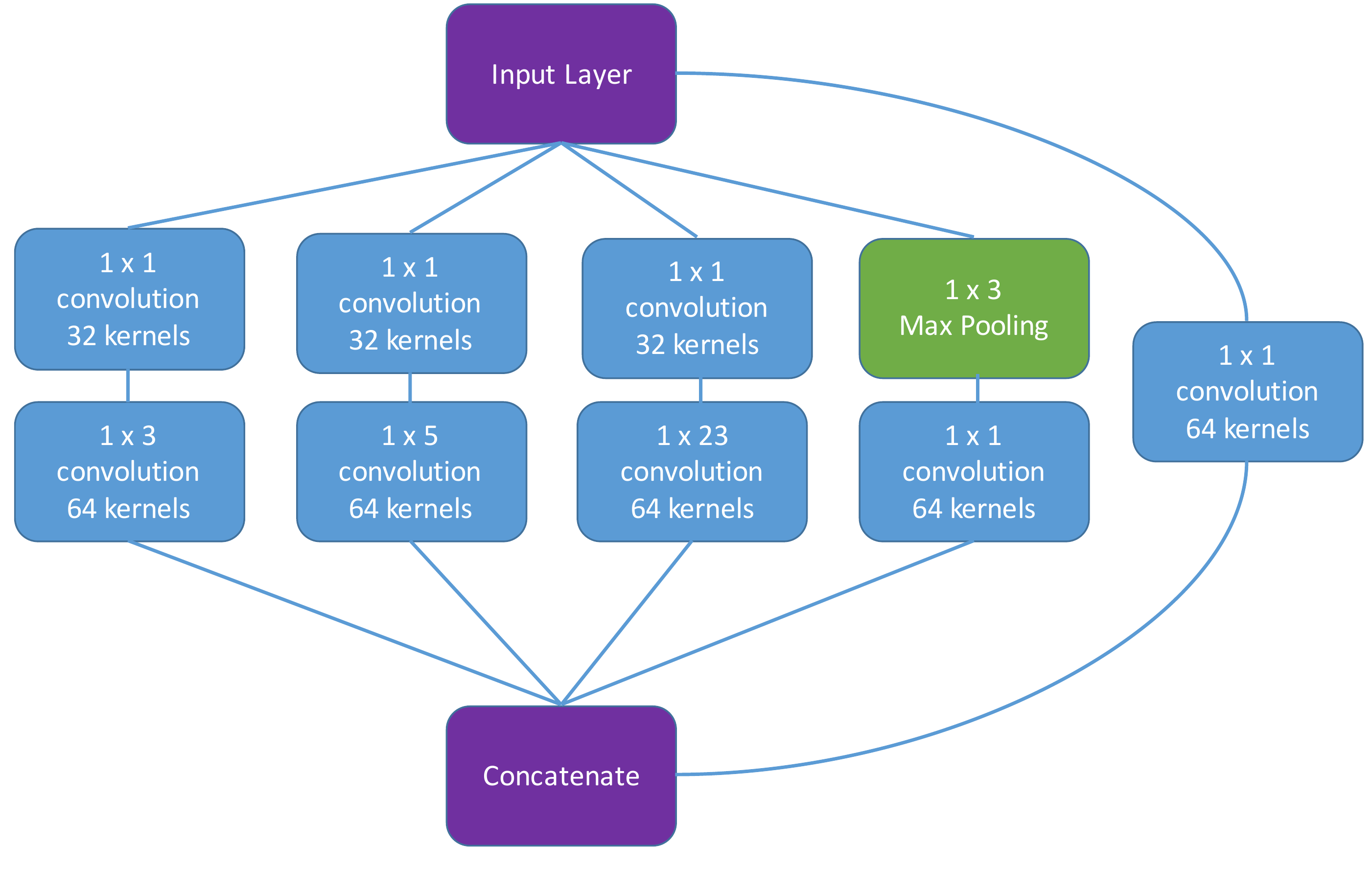}
\caption{A zoom in of an inception layer. Each of the convolutional layers (in blue) has additional batch-normalization and activation (ReLU) layers within it.} \label{inception}
\rule{9cm}{1pt}
\end{center}
\end{figure}

We considered and tested several convolutional neural networks, and the current best network has three convolutional inception layers, along with two fully-connected (dense) layers. Each inception layer has three parallel channels of convolutions, with kernel sizes of 3, 5, and 23 channels (in Faraday space each channel is 1~rad/m$^{2}$), as well as two channels with a ``1x1" convolutions, one of which also has a maxpooling layer \citep{bour10}. The 1x1 convolution essentially allows mixing between the real and imaginary parts of the spectrum in the first application, and serve to reduce the number of parameters in the two subsequent layers \citep{szeg14}. The full network is show in Fig. \ref{network}, and the individual inception layers are shown in Fig. \ref{inception}. The ``Flatten" layer will take the deep network of features constructed by the inception layers and project it into a vector of features to be passed to the dense layers. These dense layers are traditional artificial neural networks \citep{cybe89}. Source code can be found on Github\footnote{\url{https://github.com/sheabrown/faraday_complexity/blob/master/final/How2Guide.ipynb}}. 

\begin{figure}
\begin{center}
\includegraphics[width=8cm]{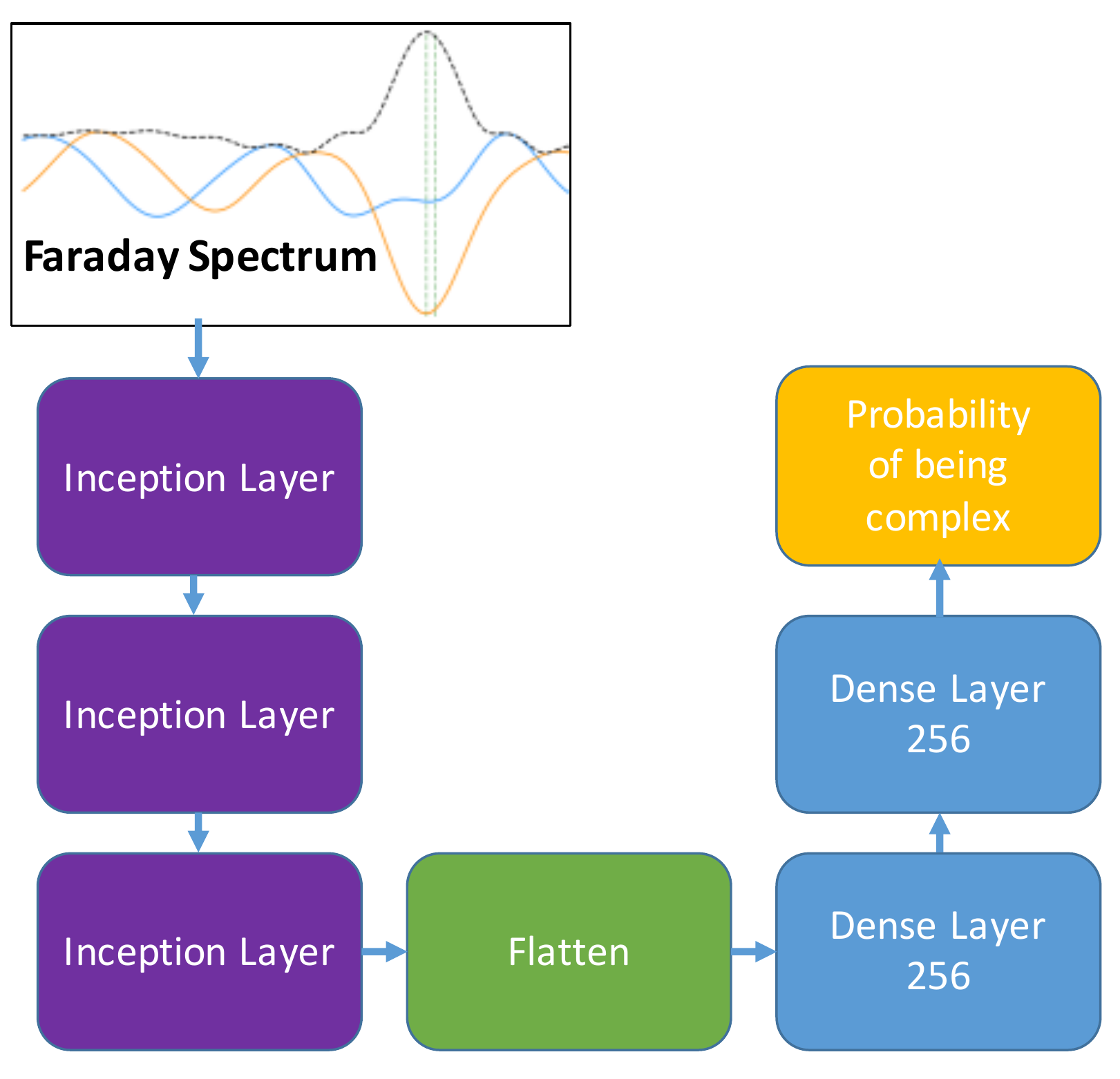}
\caption{\label{network} Three inception layer convolutional neural network (CNN) classifier. Each dense layer has an additional dropout (0.5) and activation (ReLU) layer within it.} \label{network}
\rule{9cm}{1pt}
\end{center}
\end{figure}

\section{Training} In order to train the proposed complexity classifier, we simulated both simple and complex spectra using a realistic observational model. Since our current purpose is to develop a metric for the POSSUM Early Science observations, we have used the proposed frequency coverage of the ASKAP 12 Early Science survey (700-1300~MHz, 1500-1800~MHz), with a total of 900, 1~MHz channels (see Fig. \ref{freq}). The current work focuses only on two-component models (with both components being Faraday thin), which are believed to be the dominant source of complex spectra \citep[e.g.,][]{ande15,osul17}. The two emitting regions can have different polarized amplitudes ($P_1$ and $P_2$), as well as different intrinsic polarization angles ($\chi_1$ and $\chi_2$) and foreground Faraday depths ($\phi_1$ and $\phi_2$). The combined polarized spectrum is given by  

\begin{equation} P(\lambda^2)=P_1 e^{\left[2i(\chi_1+\phi_1\lambda^2)\right]}+P_2 e^{\left[2i(\chi_2+\phi_2\lambda^2)\right]}. 
\end{equation}

\noindent We also simulate simple sources using Equation 4, but with $P_2 = 0$.  To train the network, we generated 130,000 sources (100,000 training set and 30,000 for validation), roughly half of which were complex (two-component) sources, and the other half were simple\footnote{Each time the data simulator created a spectrum, there was a 50\% probability that it would be complex. The probability is an adjustable parameter in the source code.}. Table \ref{para} shows the parameter space that was sampled at random from a uniform distribution. We used the simplifying assumption that both the noise and total intensity have no spectral dependence.  As a first step, we chose to train the network on the Faraday spectra $F(\phi)$ only, as it will be computed as part of the POSSUM pipeline, but in theory the network can be trained using the polarization spectrum directly as well.  For each source, a polarization spectrum was created first, and then a Faraday spectrum was created using the standard inversion formula of \cite{bren05}, 

 \begin{figure}
\begin{center}
\includegraphics[width=8cm]{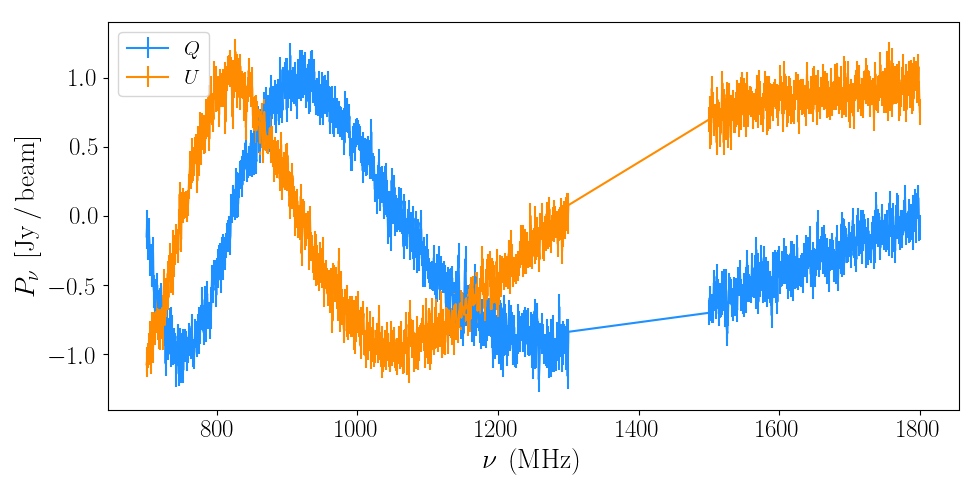}
\caption{\label{freq} An example of a polarized spectrum sampled with the POSSUM Early Science frequency coverage used in the training data. } \label{freq}
\rule{8cm}{1pt}
\end{center}
\end{figure}

\begin{equation}
F(\phi) \approx K \sum_{i=1}^{N}P_i\mathrm{e}^{-2\mathrm{i}\phi (\lambda^2_i-\lambda_0^2)},
\end{equation}

\noindent where $K=1/N_{channels}$, $\lambda_0^2$ is the average $\lambda^2$ of the channels, and $P_i$ is the measured complex polarization in channel $i$.  
 Figure \ref{freq} shows an example of a complex polarized $P(\nu)$, and Fig. \ref{complex} shows a selection of Faraday $F(\phi)$ spectra for complex sources in the training set. 
 
 The network was trained using batch stochastic gradient-descent \citep{duda12} on the training set of 100,000 sources, with 30,000 sources withheld for cross-validation during the training. The training lasted for 100 epochs, though no improvement on the validation set was found after 55 epochs. The weights found on epoch 55 were saved and used for testing. 

\begin{table} 
\vskip 0.3in
\centering
\caption{Two-component Parameter Space} \label{para}
\begin{tabular}{@{}lcc}
\hline
Parameter & Symbol & Range \\
\hline
\hline
Amplitude 1 & $P_{1}$ & 1  \\
Amplitude 2 & $P_{2}$ & [0, 1]   \\
Faraday depth \{1, 2\} & $\phi_{\{1, 2\}}$ & [-50, +50]   \\ 
Polarization angle \{1, 2\} & $\chi_{\{1, 2\}}$ & [0, +$\pi$]  \\
Noise/Channel & $\sigma$ & [0, 0.333] \\
\hline
\hline 
\end{tabular}
\end{table}

 \begin{figure*}
\begin{center}
\includegraphics[width=15cm]{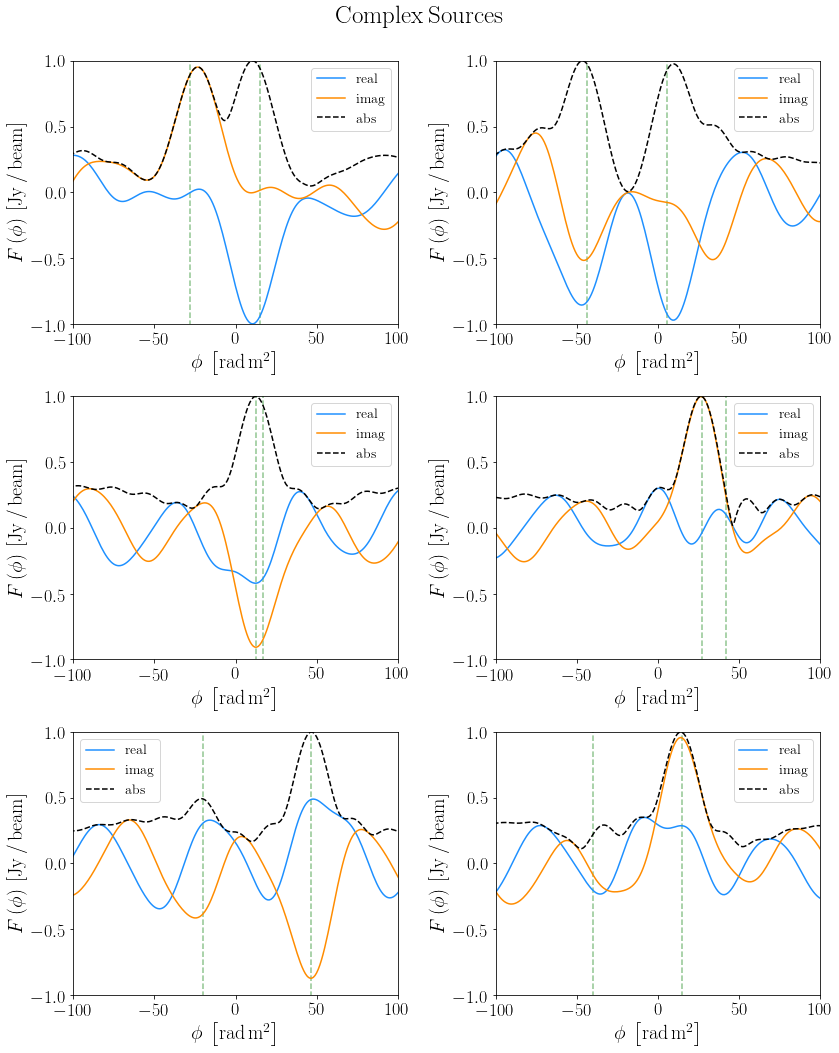}
\caption{\label{complex} Complex Faraday spectra from the training data (with $\phi_1$ and $\phi_2$ given by the vertical green dashed lines). No deconvolution (RM Clean) was performed. } \label{complex}
\rule{15cm}{1pt}
\end{center}
\end{figure*}

\begin{figure}
\begin{center}
\includegraphics[width=8cm]{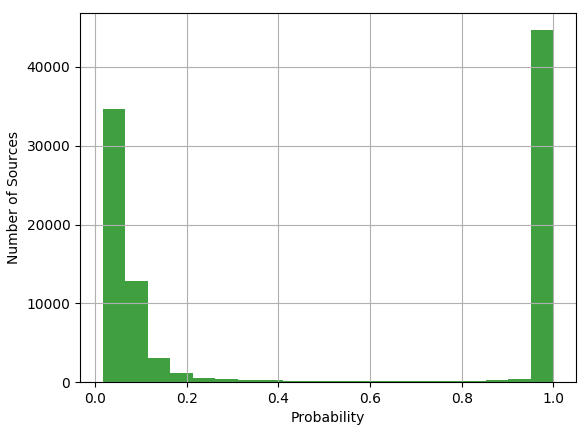}
\caption{\label{hist} Histogram of probability values for the 100,000 test sources. } \label{hist}
\rule{9cm}{1pt}
\end{center}
\end{figure}

\section{Results \& Discussion} The trained network was then applied to 100,000 new sources randomly generated using the same parameter space as the training set. The output for each source is a value $p$ between 0 and 1 which can be thought of as the probability that the source is complex. Figure \ref{hist} shows a histogram of $p$ for the 100,000 test sources. The distribution is bi-modal, indicating that the network was confident about the classification of most sources. If we take $p>0.5$ as the threshold for complexity, we can construct the confusion matrix as shown in Table \ref{conf}. 

\begin{figure}
\begin{center}
\includegraphics[width=8cm]{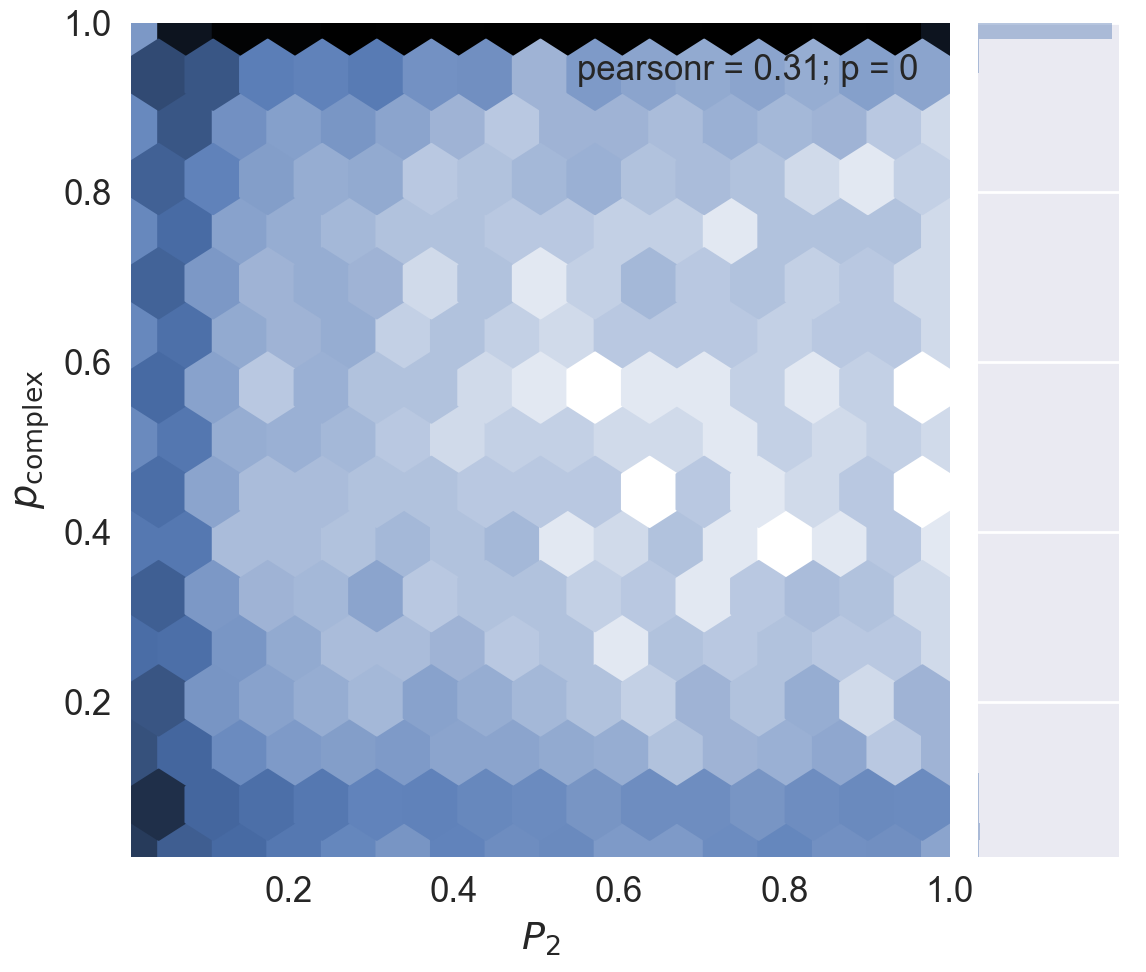}
\includegraphics[width=8cm]{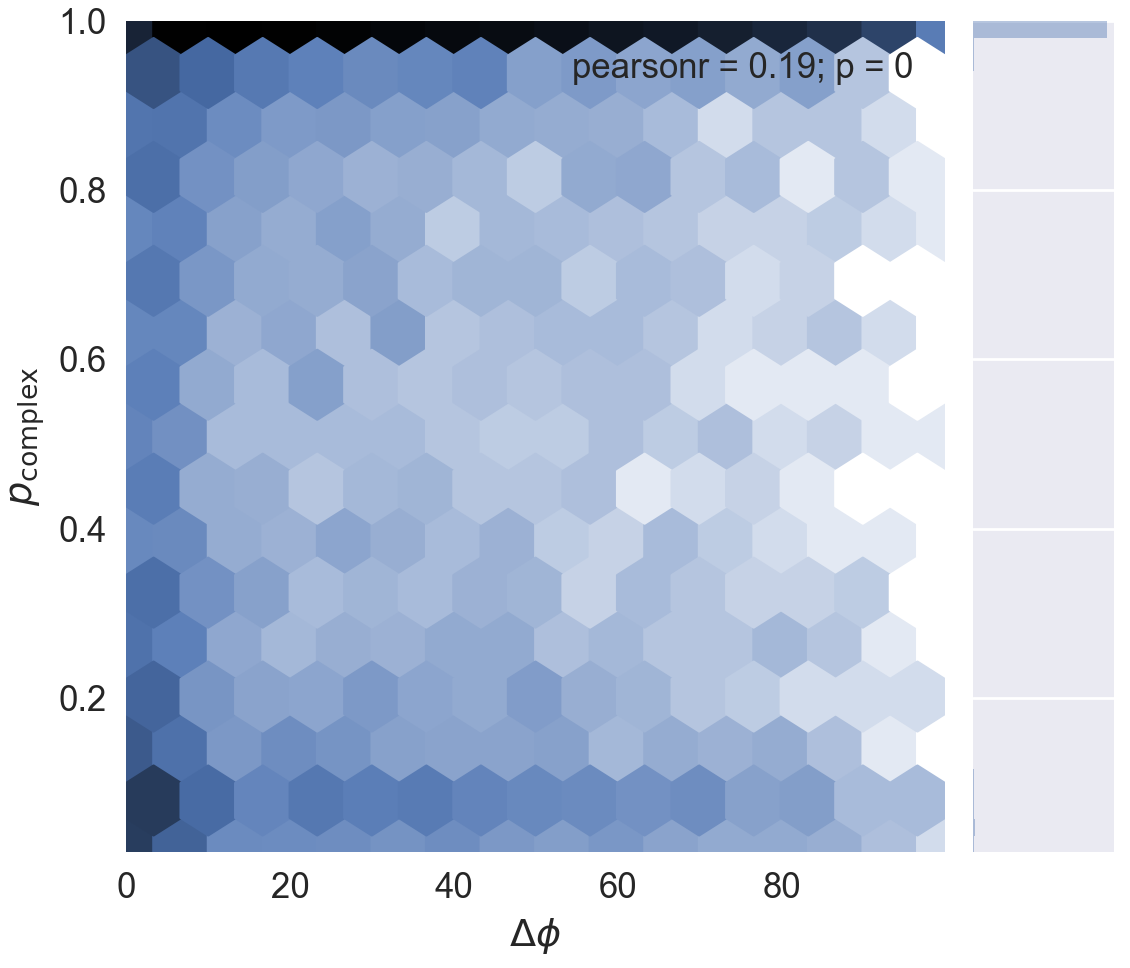}
\caption{Top: Probability output of the network vs. the relative flux of the second component for simulated complex sources in the test data set. Colour scale is the log of the number of sources. The black triangles at the top are the high density of $p$ values around 0.99, and the right sub-plot is a histogram showing that the vast majority of sources were classified as complex. Bottom: Like above, but plotted vs. the absolute difference in the Faraday depth ($\Delta \phi$) between the two components.  } \label{pv}
\rule{9cm}{1pt}
\end{center}
\end{figure}

The network produces 7.2\% false negatives and 3.0\% false positives. In order to hunt down the complex sources that are mis-classified as simple, we can plot the $p$ for all the complex sources as a function of both the second component's amplitude $P_2$ and the absolute separation in the two components' Faraday depths $\Delta \phi = |\phi_1-\phi_2|$ (Fig. \ref{pv}). The majority of false negatives happen at small $P_2$ and $\Delta \phi$. This is consistent with the results of \cite{farn11} and \cite{sun15} that point to the difficulty in identifying two component sources when $\Delta \phi < FWHM$ of the Faraday Point Spread Function. Figure \ref{missed} shows an example of one of the false-negatives. 

For the purposes of constructing a classifier for large-scale polarsation surveys like POSSUM, we would like to exclude the phase space of sources that would likely not make it into the catalog due to low signal-to-noise, as well as sources where the rotation measures of the two components are close enough to allow probing of a foreground Faraday screen. We therefore searched for the region of phase space in which we can detect $>$99\% of the complex sources, allowing for the false positive rate to adjust appropriately based on the cut-off values. We found that if the minimum signal-to-noise of the primary component is 5.0, and restrict our sample to $P_2 > 0.1$, and $\Delta \phi  >$ 2.3~rad/m$^{2}$ (which is about 10\% of the 23 rad/m$^2$ FWHM of the Faraday point spread function), the false negatives are reduced to $<$1\%, while the false positives rate reduces to $<$0.3\%. Table \ref{nconf} shows the new confusion matrix with the cut-offs applied to the same simulated data.  What these cutoffs mean for the initial POSSUM catalog (the PPC) is that the network is 99\% confident that the Faraday spectrum is simple, with the understanding that a secondary component can be hiding in the above phase space. The probability returned by the network can be recorded for each source, allowing one to flag sources where $p$ is close to the nominal cutoff of $p<0.5$ for the PPC. We should note that one can trade a higher S/N cutoff to allow a narrower $\Delta \phi$ and still reach the 99\%, something that might be advantageous depending on the science goal.  

\begin{table} 
\vskip 0.3in
\caption{Confusion Matrix: Before Cutoffs} \label{conf}
\begin{tabular}{@{}lcc}
\hline
Predicted $->$ & Simple & Complex \\
\hline
\hline
True Simple  & 48,318 & 1481  \\
True Complex & 3,618 & 46,583   \\
\hline
\hline 
\end{tabular}
\end{table}

\begin{table} 
\vskip 0.3in
\caption{Confusion Matrix: After Cutoffs} \label{nconf}
\begin{tabular}{@{}lcc}
\hline
Predicted $->$ & Simple & Complex \\
\hline
\hline
True Simple  & 29,281 & 69  \\
True Complex & 247 & 25,337   \\
\hline
\hline 
\end{tabular}
\end{table}

\begin{figure}
\begin{center}
\includegraphics[width=8cm]{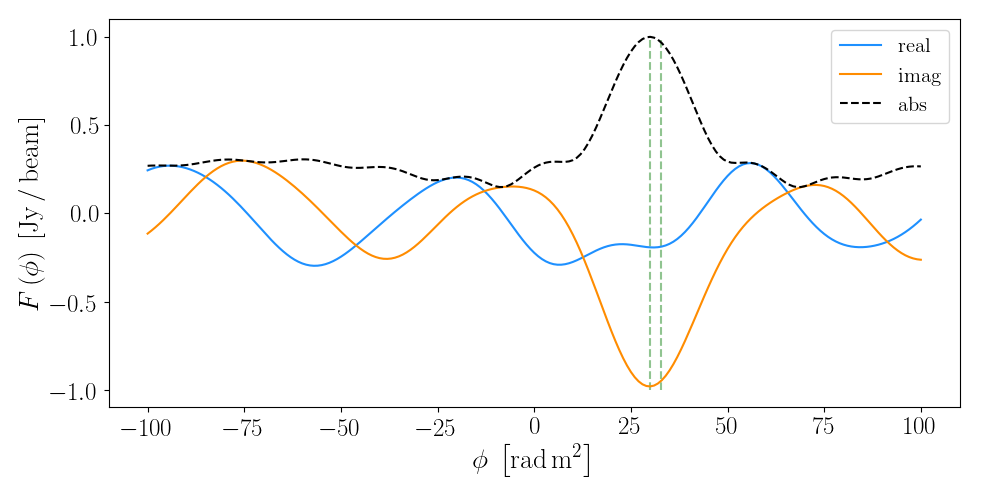}
\caption{\label{missed} Top: Faraday spectrum of a complex source mis-identified as simple by the classifier. The two Faraday depths are labeled with vertical green dashed lines. } \label{missed}
\rule{9cm}{1pt}
\end{center}
\end{figure}

\subsection{Conclusion \& Future Work} We have constructed a convolutional neural network that is able to distinguish between simple Faraday sources and those that contain two Faraday thin components, demonstrating on simulated POSSUM Early Science data that it can detect 99\% of complex sources with $<$0.3\% false positive rate in a realistic and useful region of the source parameter phase-space. The training and application of this network for other observational parameters in straightforward, needing only the frequency coverage to be changed. 
The most obvious future development of the network would include 1) lifting the simplification on the flat spectral index and channel independent noise, 2) allowing for modified RM Synthesis that includes channel weights in Equation \ref{inversion}, and 3) the inclusion of Faraday thick and three component sources during training. Including complexity beyond this may prove impractical, as \cite{osul17} was able to fit just about any source using a combination of three Faraday thin components. Given the power of CNNs used in commercial applications, the inclusion of 3) in to the training would also allow for multiple classes beyond a binary simple/complex classification.  

The Dunlap Institute is funded through an endowment established by the David Dunlap family and the University of Toronto. B.M.G. acknowledges the support of the Natural Sciences and Engineering Research Council of Canada (NSERC) through grant RGPIN-2015-05948, and of the Canada Research Chairs program. Partial support for LR comes from NSF grant 17-14205 to the University of Minnesota. XHS is supported by the National Natural Science Foundation of China under grant No. 11763008.












\bsp	
\label{lastpage}
\end{document}